# Career Mobility of Planning Alumni in the United States: Evidence from Professional Profile Data Using Large Language Models

**Yan Wang**[1*] and **Su Jeong Jo**[1]

[1] *Department of Urban and Regional Planning & Florida Institute for Built Environment Resilience, University of Florida, Gainesville, FL 32611, USA*

* Correspondence: yanw@ufl.edu | ORCID: 0000-0002-3946-9418

Su Jeong Jo: sujeongjo@ufl.edu | ORCID: 0009-0001-6860-3866

**Published in:** *Journal of the American Planning Association* | DOI: 10.1080/01944363.2026.2668445

**Abstract**

**Problem, Research Strategy, and Findings —** Planning professions in the United States navigate complex and dynamic career landscapes under rapid urban changes, yet comprehensive evidence regarding their career trajectories, advancement patterns, and the influence of social, spatial, organizational, and educational factors remains limited. This study draws on boundaryless career theory, social capital theory, and spatial opportunity models to analyze career mobility among more than 130,000 planning alumni. Using large language models (LLMs) to extract structured information from LinkedIn profiles, our results reveal that planning alumni who adopt boundaryless career patterns, specifically multisector experience or lateral and industry-switching trajectories, achieve significantly higher upward mobility. While technical competencies provide a foundational entry-level signal, soft skills leveraged through strategic lateral moves become increasingly decisive as planners reach senior stages. Geographic mobility and employment in larger, diverse metropolitan labor markets are both associated with advancement, though the latter provides modest benefits. Larger professional networks and greater organizational engagement are consistently associated with upward career transitions, while AI-related skills, now commonplace, present limited additional advantage. Limitations include reliance on LinkedIn data, which may underrepresent alumni without online profiles, and an individual-level focus that omits organizational factors.

**Takeaway for Practice —** For planning alumni, cultivating broad sector experiences, expanding professional networks, seeking geographic mobility, and prioritizing ongoing technical and organizational learning are key advancement strategies. These results have direct implications for planning educators, employers, and policymakers designing programs to support an adaptive, well-connected, and effective workforce.

## 1. INTRODUCTION

Planning in the United States has always been a profession in motion [1]. Planners navigate across public, private, and nonprofit sectors; shift between technical, managerial, and policy-oriented roles; and relocate among diverse regions [2]. As cities and regions face rapid technological change, housing crises, climate imperatives, and renewed calls for equity [3], the planning profession is under pressure to adapt—not only in what it plans for, but in how planners themselves build and sustain their careers.

Career mobility, defined as the movement of professionals across roles, organizations, sectors, and locations over time [4, 5], stands at the center of this adaptive challenge. Foundational theories from social stratification and organizational sociology emphasize that upward mobility depends not only on personal credentials, but also on opportunity structures, institutional pathways, and accumulated career capital, spanning technical expertise, social networks, and reputational standing [6, 4, 7].

Contemporary analyses highlight the rise of "boundaryless" and "protean" careers, in which professions like planning increasingly make lateral moves, cross-sector transitions, and self-directed searches for meaningful opportunities rather than orderly progression up traditional hierarchies [8, 9]. Understanding the mechanisms and patterns of career mobility among planning alumni is thus essential both for supporting individual career resilience and for ensuring that planning education, workforce development, and professional institutions remain responsive to the rapidly evolving demands of urban and regional governance.

Traditional research on planning careers has relied on alumni or workforce surveys that offer useful, if limited, snapshots. McClure et al. [10] found that while planning education exposes students to diverse theoretical perspectives, early professional experience often shifts graduates toward pragmatic identities. Guzzetta and Bollens [11] identified specialized competencies distinguishing planners across sectors and career stages. Keyes and Ames Fischer [12] addressed undergraduate planning education within a Hispanic-Serving Institution. However, these studies are typically limited by small sample sizes or cross-sectional designs that cannot capture the longitudinal flow of careers over time.

The recent explosion of professional networking platform data such as LinkedIn has enabled a new wave of empirical career-mobility studies across many fields [13, 14, 15, 16, 17, 18]. Despite this methodological progress, systematic evidence on the career mobility of planning alumni remains limited, especially concerning the interplay of spatial context, cross-sector transitions, professional networks, and technical skill development. In this study, we leverage recent advances in large language models and statistical models to rigorously annotate and analyze professional profile data from planning alumni at scale. By integrating perspectives from sociology, geography, organizational studies, and labor economics, we seek to answer: (1) How do sectoral backgrounds, spatial contexts, network characteristics, and skill acquisition shape career advancement among planning alumni? (2) What mechanisms and opportunity structures are associated with upward career mobility in the planning profession?

## 2. LITERATURE REVIEW

### 2.1 Sociological and Interdisciplinary Foundations of Career Mobility

Career mobility theory is rooted in sociology, specifically social stratification, labor studies, and organizational sociology [4, 7]. Economists study career mobility through wage growth, job changes, and the impacts of education and skills; organizational behavior scholars apply mobility theory to understand promotions, leadership pipelines, and organizational design; and psychologists examine motivation, personality, and career decision-making [19, 5]. This diversity underscores that careers unfold at the intersection of personal agency and structural constraint.

Within these frameworks, careers are understood as sequences of job transitions and positional changes over working life, shaped by both personal agency and structural context [4, 5]. Careers differ in the amount of mobility, the direction of job transitions—whether they involve upward or downward changes in earnings, status, or noneconomic rewards—and their orderliness (the extent to which successive jobs share characteristics such as skills) [5]. Career mobility depends on both personal attributes and contextual factors [20, 21]. In planning, mobility can be further shaped by institutional accreditation, professional licensing, and fluctuating spatial demands across metropolitan and rural contexts [22, 23, 24].

### 2.2 Sectoral and International Experiences

Traditional models—internal labor market theory [25] and occupational labor markets [7]—emphasize hierarchical progression within a single organization. However, contemporary careers in knowledge-intensive professions like urban planning increasingly follow boundaryless and protean patterns, characterized by lateral moves, sectoral transitions, geographic relocation, and self-directed skill acquisition [8, 9]. Central to understanding these trajectories is the concept of career capital [6]: human capital (knowledge, education, technical skills), social capital (professional connections and reputational networks), and reputational capital (credibility within the field).

Research indicates that exposure to diverse sectoral experiences is linked to upward mobility and entry into prestigious organizations [13, 14]. Portfolio or non-linear career paths can accelerate leadership attainment compared to traditional linear advancement [8, 9], and international education has been positively associated with mobility across sectors and locations [15]. These dynamics are salient in planning, where career paths often cross public, private, and nonprofit boundaries due to the field's dual accountability to both public interest and market efficiency [23, 24]. We hypothesize:

> ***Hypothesis H1:*** *Planning alumni with multisector experience (public and private sectors), broader industry exposure span, boundaryless career trajectories (portfolio, industry switch, or lateral), or international educational experience demonstrate faster and greater upward role advancement, controlling for career length, education, and AICP credential.*

## 2.3 Spatial Contexts and Geographical Mobility

Spatial context is a critical but underexamined determinant of career mobility in planning. Research in geography and urban studies has demonstrated that city size, labor market diversity, and regional accessibility are linked to upward and lateral mobility for skilled workers [26, 27]. Abundant studies from the intergenerational mobility literature reveal the space and place effect: residential environment has a causal impact on economic and social mobility [28]. For example, Michelangeli and Türk [29] show that intergenerational upward mobility is higher in larger, more accessible cities, while Giorgi et al. [30] demonstrate that interregional mobility of highly skilled workers fuels technological novelty. Conversely, spatial mismatch can limit advancement [31].

This spatial dimension is fundamental to urban planning, as practitioners' work is intrinsically tied to place and fluctuating demands across cities, regions, and metropolitan markets [23, 24]. Planning contexts are heavily shaped by local governmental structures and regional policy priorities [32]. We therefore hypothesize:

> ***Hypothesis H2:*** *The high market diversity of large metropolitan regions facilitates planning alumni's geographic, sectoral, and industrial mobility, thereby amplifying the positive effects of non-linear career trajectories on career advancement, controlling for career length, education, and AICP credential.*

## 2.4 Social Network Effects

Social capital theory posits that access to broad professional connections, mentoring, and reputational networks accelerates career mobility by providing information, referrals, and support [33, 34]. Recent work shows that broader and more heterogeneous networks facilitate innovation and speed career progression across various occupations [35, 36]. In planning, career advancement often depends on building professional identity and trust through networks, as the field is inherently shaped by public trust, political context, and inter-organizational collaboration [37]. Despite these established professional values, empirical evidence linking specific network affiliations to quantifiable mobility outcomes remains limited. We hypothesize:

> ***Hypothesis H3:*** *Alumni with larger professional networks and more organizational engagement experience faster and greater career mobility over time regardless of career length, career stage, education, and AICP credential.*

## 2.5 Technical Skills and Continuing Learning

Accumulating new technical skills and credentials through certifications, formal coursework, or self-directed development represents a key driver of ongoing employability and upward mobility [38]. Within the planning field, continuous learning increasingly encompasses urban analytics, geographic information systems, project management, and AI/machine learning [39, 40, 41]. The rapid normalization of core digital skills raises questions about which technical proficiencies continue to differentiate candidates and drive advancement. We therefore hypothesize:

> ***Hypothesis H4:*** *Planning alumni who continuously acquire new skills (soft and technical) and certifications will experience faster and greater upward transitions, with this effect differing significantly across career trajectories, controlling for career length, education, and AICP credential.*

## 3. DATA AND METHODS

We compiled a dataset of professionals with an urban planning degree (planning alumni hereafter) using LinkedIn People Profiles (Bright Data, 2025) within the U.S. Each record includes career history, educational background, geographic location, skills, and professional network information. Data were filtered to include only profiles with a verified urban planning degree and valid, trackable post-graduation professional employment history. All analyses used publicly available, non-private profile information, and the data obtained from Bright Data were de-identified. The study protocol was reviewed to comply with institutional and publisher ethics requirements.

### 3.1 Data Collection and Preprocessing

Our raw dataset, retrieved on August 27, 2025, comprised 154,553 LinkedIn profiles of planning alumni identified via degree-relevant keywords derived from Planning Accreditation Board (PAB) and Association of Collegiate Schools of Planning (ACSP) accredited programs. To ensure data quality, we excluded profiles lacking professional experience, sufficient date information, or those listing only a single position. We applied the U.S. Social Security Administration's retirement age of 66 as a threshold to exclude implausible career durations. Following preprocessing, 133,888 profiles remained.

Geographical mobility—including relocations and urban-rural shifts—was tracked across each individual's career by classifying locations at the state, urban, and metropolitan levels based on 2020 OMB definitions and TIGER/Line shapefiles [42]. Market exposure (cumulative industrial diversity over a career) was computed as a duration-adjusted average of labor market diversity, weighted by months spent in each location. Using 2023 BLS Quarterly Census of Employment and Wages data [43], we measured industry evenness within each CBSA via Pielou's evenness index (J) [44], where 1 indicates perfect diversification and 0 indicates industry dominance. Skills were extracted from "courses," "certifications," and "experience" fields, then mapped to O*NET taxonomies for Urban and Regional Planners (19-3051.00) [45].

### 3.2 Classifying Career Trajectory with Large Language Models

To process semi-structured LinkedIn data—including inconsistent job titles and noisy narrative text [46]—two LLMs, GPT-4o-mini and o4-mini, were employed for cross-model validation. This approach facilitated the large-scale transformation of unstructured career histories into structured annotations. Seven labels were predefined for annotation at job or profile levels: Sector, Seniority, Industry Type, Planning-related or not, Trajectory Type, International Education Experience, and Career Mobility.

While rule-based methods struggle with the contextual interpretation required for professional language [47], these LLMs were selected for their superior reasoning capabilities based on publicly available benchmark rankings [48]. Given the trade-off between output reliability and token limits [49], 2.6% of exceptionally dense profiles were excluded. Both models were accessed through the University of Florida's NaviGator APIs, yielding 130,301 profiles for analysis. Detailed prompts to mitigate hallucination and rigorous validation metrics are provided in Appendix B.

### 3.3 Statistical Design for Hypothesis Testing

To analyze upward mobility across two dimensions—frequency (count) and timing (speed)—Negative Binomial (NB) Regression and Cox Proportional Hazards (Cox PH) models were employed. The NB model addressed overdispersion in advancement counts, while the Cox PH model captured the temporal dynamics of career progression. Models were controlled for Career Length, Education Level, and AICP Certification. AICP certification is a primary professional signal associated with a 31% salary premium [50] and was included to isolate its influence from the independent variables.

A multi-staged approach accounted for the non-linear, heterogeneous nature of career paths. Beyond the full models (Main and Detailed models in Table 2), Quartile Models (Q1–Q4) were estimated by career tenure to capture evolving mobility rewards. Variables and interaction terms were iteratively incorporated, monitoring model fitness (AIC, Pseudo $R^2$ for NB; C-index for Cox PH).

**Table 1. Key Variables, Definitions, and Measures**

| Category | Variable | Description & Measurement | Data Type |
|---|---|---|---|
| **Social Capital** | Connections | Total LinkedIn professional connections | Discrete |
| | Organizations | Total unique non-occupational affiliations | Discrete |
| **Career History** | Career Length | Total tenure in months (excluding gaps) | Continuous |
| | Stability | Average tenure per job | Continuous |
| **Geographical Mobility** | Total Moves | Cumulative count of geographical relocations | Discrete |
| | State Moves | Relocations across state borders | Discrete |
| | Intl Moves | Relocations across international borders | Discrete |
| | Urban-Rural Shift | Transitions between urban and rural areas | Discrete |
| **Market Context** | Shannon | Market diversity; duration-weighted CBSA-level Shannon entropy (2-digit NAICS) | Continuous |
| | Population | City size; duration-weighted CBSA-level population | Continuous |
| **Human Capital** | AICP Flag | Professional planning certification | Binary |
| | Total Skills | Count of all skills extracted from profile | Discrete |
| | Soft Skills | Number of soft skills | Discrete |
| | Tech Skills | Number of technical skills | Discrete |
| | AI Skills | Number of AI-related skills | Discrete |
| | Edu Level | Highest degree: Bachelor's, Master's, Doctoral | Categorical |
| | Planning Exit | Planning as terminal degree | Binary |
| **Boundaryless Career (LLM)** | Multisector | Sectoral breadth: worked in public and private sectors | Binary |
| | Academia | Experience in academia | Binary |
| | Industry Span | Count of distinct 2-digit NAICS industry sectors | Discrete |
| | Trajectory Types | Linear, Lateral, Portfolio, Industry Switch | Categorical |
| | Boundaryless | Experience in portfolio or industry switch trajectory | Binary |
| | International | International education experience | Binary |
| **Outcome (Y)** | Upward Mobility | Count of transitions to higher positions, responsibility, or authority | Discrete |

**Table 2. Statistical Design**

| Model | Independent Variables (Interaction Focus) | Controls | Outcome |
|---|---|---|---|
| **Main (NB & Cox PH)** | Multisector, Academia, Industry Span, Boundaryless, International, Shannon, Population, Total Moves, Connections, Organizations, AICP, Total Skills, Edu Level, Planning Exit, Stability. Interactions: Boundaryless x Career Length, x Connections, x Multisector, x Industry Span | Career Length, Edu Level, AICP Flag | Upward Transitions (Count & Time-to-Event) |
| **Detailed (NB & Cox PH)** | All main variables plus: Trajectory Types (Portfolio, Industry Switch, Lateral), State/Intl Moves, Urban/Rural Shifts, Soft/Tech/AI Skills. Trajectory x Skill, x Move, x Shift interactions | Career Length, Edu Level, AICP Flag | Upward Transitions (Count & Time-to-Event) |

**Table 3. Negative Binomial Regression Results (Main Model)**

| Variable | Coeff. | IRR | 95% CI |
|---|---|---|---|
| Intercept | 0.01 | 1.01 | 0.98–1.04 |
| C(Career Stage Q2)*** | 0.27 | 1.31 | 1.28–1.33 |
| C(Career Stage Q3)*** | 0.29 | 1.33 | 1.31–1.36 |
| C(Career Stage Q4)*** | 0.19 | 1.21 | 1.18–1.24 |
| Multisector*** | 0.10 | 1.10 | 1.08–1.12 |
| Industry Span*** | 0.05 | 1.06 | 1.05–1.06 |
| Academia*** | 0.11 | 1.12 | 1.10–1.14 |
| International Education* | 0.02 | 1.02 | 1.00–1.04 |
| Boundaryless*** | −0.38 | 0.68 | 0.67–0.69 |
| Shannon | 0.01 | 1.01 | 1.00–1.02 |
| Population*** | 0.01 | 1.01 | 1.01–1.01 |
| Total Moves*** | 0.06 | 1.06 | 1.06–1.06 |
| Connections*** | 0.0009 | 1.0009 | 1.0008–1.0009 |
| Stability*** | 1.01 | 1.00001 | 1.000009–1.00001 |
| Organizations*** | 0.04 | 1.04 | 1.02–1.05 |
| Total Skills*** | 0.01 | 1.01 | 1.01–1.01 |
| AICP Flag | 0.04 | 1.04 | 1.00–1.08 |
| Planning Exit* | 0.02 | 1.02 | 1.002–1.05 |
| Edu (Master) | 0.02 | 1.02 | 1.00–1.05 |
| Edu (Doctoral) | −0.005 | 0.99 | 0.94–1.05 |
| Boundary x Tenure*** | 0.0003 | 1.0003 | — |
| Boundary x Connections*** | 0.0004 | 1.0004 | 1.0004–1.0005 |
| Boundary x Multisector*** | 0.07 | 1.07 | 1.05–1.10 |
| Boundary x Industry*** | 0.10 | 1.10 | 1.09–1.11 |

*Obs.: 130,301 | Pseudo $R^2$: 0.0418 | AIC: 541,089.65 | Dispersion: 0.3340 | *** $p < 0.001$, ** $p < 0.01$, * $p < 0.05$; no star $p \geq 0.05$*

**Table 4. Cox Proportional Hazards Results (Main Model)**

| Variable | Coeff. | HR | 95% CI |
|---|---|---|---|
| Multisector*** | 0.19 | 1.21 | 1.19–1.24 |
| Industry Span*** | 0.03 | 1.03 | 1.03–1.04 |
| Boundaryless*** | −0.48 | 0.62 | 0.61–0.63 |
| Academia*** | 0.08 | 1.08 | 1.07–1.10 |
| International Education*** | 0.12 | 1.13 | 1.11–1.14 |
| Shannon*** | 0.02 | 1.02 | 1.01–1.03 |
| Connections*** | 0.0007 | 1.0007 | 1.0006–1.0007 |
| Stability*** | 0.000008 | 1.000008 | 1.000007–1.000009 |
| Organizations*** | 0.07 | 1.08 | 1.06–1.09 |
| Total Skills*** | 0.01 | 1.01 | 1.01–1.01 |
| Career Length*** | −0.006 | 0.99 | 0.99–0.99 |
| AICP Flag*** | 0.06 | 1.07 | 1.03–1.10 |
| Edu (Master)*** | 0.05 | 1.05 | 1.03–1.07 |
| Edu (Doctoral)** | −0.07 | 0.93 | 0.89–0.98 |
| Boundary x Career Length*** | −0.0003 | 0.10 | 0.9996–0.9998 |
| Boundary x Connections*** | 0.0003 | 1.0003 | 1.0002–1.0003 |
| Boundary x Multisector*** | 0.10 | 1.10 | 1.08–1.13 |
| Boundary x Industry*** | 0.05 | 1.05 | 1.04–1.06 |

*Obs.: 130,299 | Events: 112,095 | AIC: 1,813,471.47 | C-index: 0.7183 | Strata: Planning Exit, Population, Total Moves | *** $p < 0.001$, ** $p < 0.01$, * $p < 0.05$*

## 4. FINDINGS

We report Incidence Rate Ratios (IRR) and Hazard Ratios (HR) from NB and Cox PH models, respectively. Full model diagnostics including estimates, p-values, and confidence intervals are detailed in Appendix D.

### 4.1 Patterns of Career Length, Job Count, and Stability

We analyzed 130,301 U.S. planning alumni profiles using quartile-based segmentation (Q1–Q4). The median career length was 15.4 years, with a range of 9.1 (Q1) to 23.7 years (Q3) (Figure 1a). This duration aligns with the 2025 APA Salary Survey (median: 11 years) and reflects the peak mobility window where significant transitions occur. Profiles averaged 6.6 jobs, with a 25th–75th percentile range of 4 to 9. Career stability—career length divided by job count—increases from a mean of 1.14 to 7.05 years per job as professionals move from early (Q1) to late career (Q4). Overall, 16.6% of profiles showed low stability (≤1 year), 62.2% medium (2–5 years), and 21.2% high stability (>5 years) (Figure 1b).

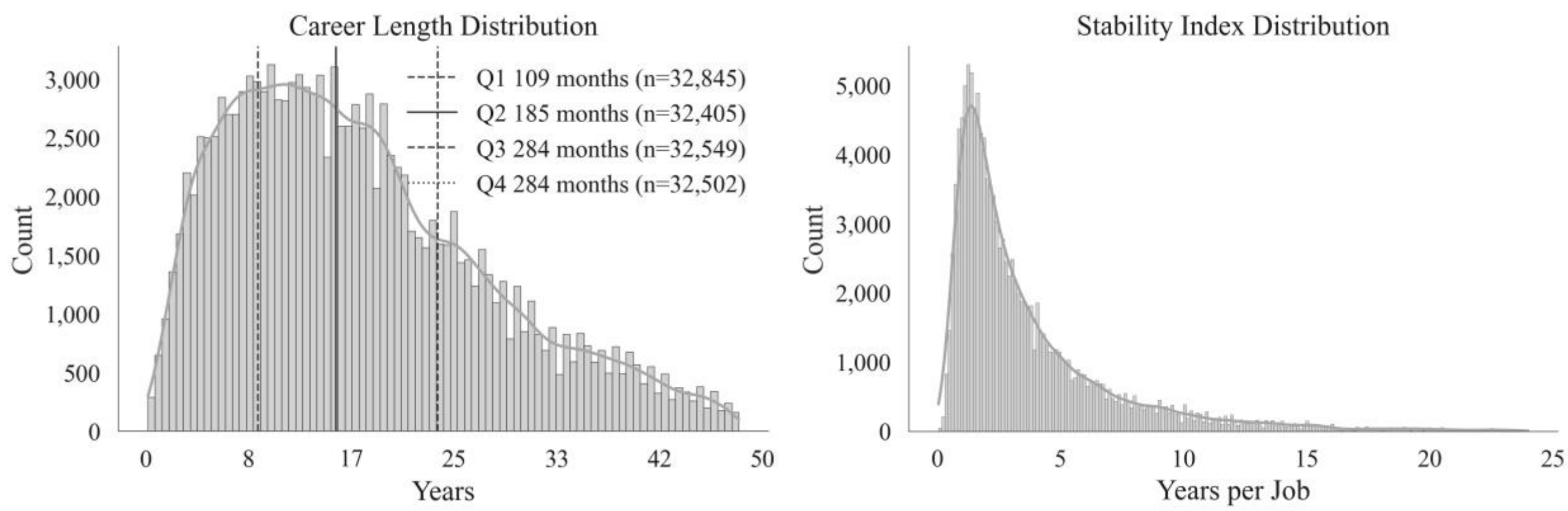

*Figure 1. Career Length Distribution (left) and Stability Index Distribution (right) by quartile. Dashed lines indicate Q1–Q4 thresholds. n = 130,301.*

## 4.2 Strategic Breadth and Trajectory: Sector, Industry, and Boundaryless Careers

The private sector (39.7%) dominates the distribution, followed by government (23.0%) and nonprofit (18.4%). Nearly half of alumni exhibit linear patterns (49.9%), while industry switching (21.4%) and lateral movement (20.0%) are prevalent. Career mobility outcomes are predominantly upward (69.5%), followed by lateral (15.7%) and stable (6.2%)—a trend potentially driven by LinkedIn's self-promotion bias.

Results from both NB and Cox PH models reveal a robust consensus: factors correlated with higher promotion frequency also align with accelerated advancement speed. A boundaryless career pattern is generally associated with a 31.9% lower promotion frequency (IRR = 0.681) and a 38.2% decrease in speed (HR = 0.618). However, this relationship is highly dynamic across career stages: in the earliest phase (Q1), boundaryless patterns show a strong positive association with advancement frequency (IRR = 2.54), indicating that early-career planning alumni leverage organizational changes as a strategic tool for role exploration.

The main and detailed models suggest a potential compensatory association with tenure, where the initial friction observed in boundary-crossing is mitigated over time: boundaryless–tenure (IRR = 1.0003), industry switch–tenure (IRR = 1.0002), and lateral–tenure (IRR = 1.0003) indicate that while frequent switching creates immediate friction, the diverse experiences acquired become increasingly valuable assets over the long term. Interaction effects for boundaryless–multisector (IRR = 1.077; HR = 1.105) and boundaryless–industry breadth (IRR = 1.09; HR = 1.05) suggest that boundaryless patterns yield positive outcomes when coupled with sectoral or industrial breadth. Individuals spanning public and private domains (IRR = 1.101; HR = 1.215) or possessing academic experience (IRR = 1.120; HR = 1.082) exhibit consistent advancement advantages. International education is linked to significantly accelerated promotion velocity (HR = 1.253), a trend particularly pronounced in later career stages (Q3 IRR = 1.037; Q4 IRR = 1.071).

## 4.3 Spatial Context and Geographic Mobility

Geographic mobility is widespread: 67.7% of alumni moved internationally, and 49.2% of those with known U.S. transitions crossed state borders at least once. Each additional interstate move is linked to a 2.4% increase in promotion frequency (IRR = 1.024) and a 4.5% acceleration in promotion speed (HR = 1.045). Population size (IRR = 1.011) is consistently associated with upward mobility, highlighting the advantages of larger labor markets. Market diversity measured by Shannon entropy (IRR = 1.009; HR = 1.045) exhibits a statistically significant link to promotion speed. Geographic mobility shows positive correlations when interacted with boundaryless patterns, such as portfolio–state moves (IRR = 1.034), industry switch–urban-to-rural shift (IRR = 1.072; HR = 1.053), and lateral–rural-to-urban shift (IRR = 1.055). Jurisdictional relocations during industrial shifts (IRR = 1.032) or lateral moves (IRR = 1.058) demonstrate broader professional adaptability linked to increased promotion frequency.

### 4.4 Social Network Effects: Relational Capital and Market Connectivity

Network expansion is most rapid during early career (Q1), averaging 58.9 new connections annually (mean: 268), before slowing in mid-career (Q2: 25.8/year; Q3: 16.7/year) and plateauing in late stage (Q4: 10.1/year; mean: 309). A clear inverse relationship exists between career stability and network size: each additional year in a single position correlates with a decline in median connections, most pronounced in early stages (Q1 slope = −5.55; Q2 slope = −4.12) (Figure 2).

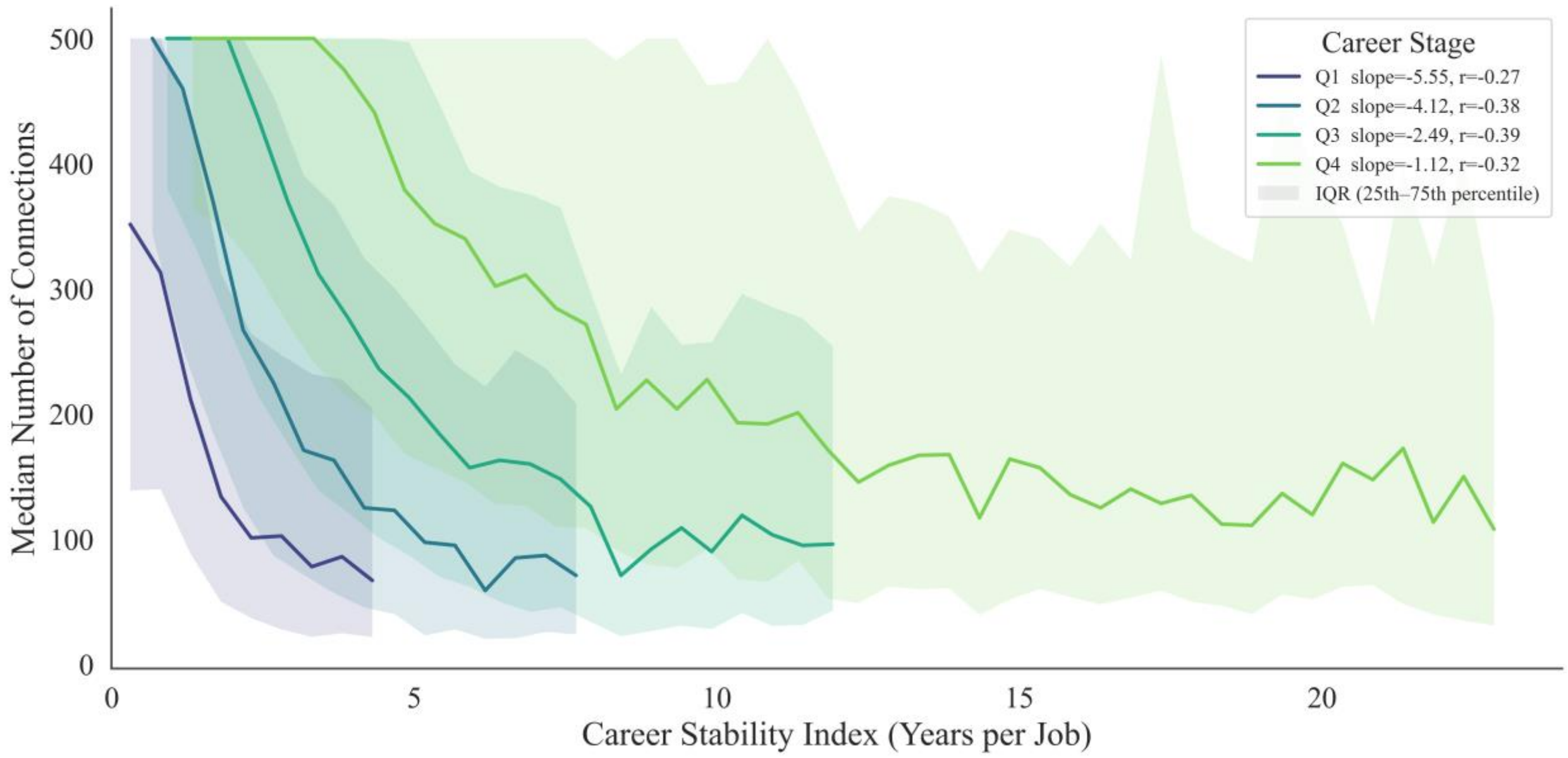

*Figure 2. Median number of LinkedIn connections by Career Stability Index across career stages (Q1–Q4). Shaded bands indicate the interquartile range (25th–75th percentile).*

Each additional connection is linked to a statistically significant increase in promotion frequency (IRR = 1.001) and promotion speed (HR = 1.0005). This relational advantage remains stable across the career span (Q1–Q4 IRR = 1.001). Total unique non-occupational affiliations remain a significant predictor of growth (IRR = 1.026; HR = 1.021), illustrating a "network-driven" mechanism in which relational capital acts as a structural facilitator.

### 4.5 Human Capital and Skill Specialization

The sample is predominantly composed of Bachelor's degree holders (83.37%), with a significant convergence of disciplinary focus at the graduate level: 79.07% of Master's degree holders maintain a planning-aligned degree. Linear upward trajectories are most prevalent among Doctorate holders (51.76%), who also exhibit the highest international exposure (29.07%). Bachelor's and Master's holders show a higher propensity for non-linear paths (46.5%), driven by a balanced mix of industry switching (21%) and lateral movement (20%).

Soft and management competencies (62.43%) are most prevalent, led by instructing (24.21%) and monitoring (15.51%). Technical skills (59.36%) remain anchored in foundational software like the Microsoft suite (23.22%) and spatial analysis tools such as GIS (16%). AI-related competencies (24.61%)—including machine learning (19.78%) and generative AI (8.83%)—currently function as a baseline requirement rather than a unique strategic advantage, with a significant negative association between AI skills and promotion frequency (IRR = 0.917).

Overall skill breadth shows a consistent positive association with growth: each additional skill is linked to a 1.3% increase in promotion frequency (IRR = 1.013; HR = 1.008). A notable trade-off emerges: soft skills demonstrate a robust positive association with promotion frequency (peaking in Q1: IRR = 1.032) but correlate negatively with promotion speed (HR = 0.987). In contrast, technical competencies function as the primary engine for career velocity (HR = 1.077), a particularly potent advantage when coupled with lateral moves (IRR = 1.067).

Traditional credentials like AICP certification (IRR = 1.077; HR = 1.050) and Master's degrees (HR = 1.073) provide stable advantages.

# 5. DISCUSSION

## 5.1 Implications for Practice and Education

Situated at the interplay of career capital theory, boundaryless career theory, spatial opportunity models, and social capital theory, this empirical study bridges proven insights from broader career mobility research with the unique, spatially embedded, and network-rich context of American planning alumni.

At the micro level, the evidence underscores the value of proactive career management. Upward mobility is strongly associated with broad networks, multi-organizational engagement, cross-sectoral mobility, and strategic relocation to diverse markets. Crucially, technical skills function as a foundational entry-level signal, while soft skills leveraged through strategic lateral moves become increasingly decisive as planners transition into senior roles—aligning with Barchers et al. [51] on the primacy of leadership in senior roles. Mentorship initiatives should encourage early-career alumni to deliberately diversify their professional experiences.

At the organizational and educational level, planning programs should look beyond traditional employment outcomes and align training with emerging professional upskilling priorities. This includes strengthening curricula with technical instruction, embedding training in network-building, professional organization participation, and career navigation skills. Workforce development efforts would benefit from supporting planners in acquiring transferable expertise and fostering connections across public, private, and nonprofit sectors.

At the field level, planning alumni's career advancement is not driven by a single linear pathway but instead reflects how professional capital accumulates across sectors, organizations, networks, and spatial contexts. Policies and educational practices that reduce barriers to mobility and network participation while cultivating broad skill sets may yield a more effective and equitable planning profession.

## 5.2 Limitations and Opportunities

This study has several limitations inherent in its reliance on LinkedIn data. The sample may underrepresent planners with limited digital presence or non-traditional educational pathways. As self-reported data, profiles may reflect self-presentation bias: career advancement and digital visibility often form a self-reinforcing cycle. LLM-based annotation, which relies on semantic reasoning, may inadvertently amplify these self-presentation effects.

The observational structure presents inherent measurement challenges. Recorded job counts may include volunteer or project-based roles, potentially inflating transition metrics. The observed associations should not be interpreted causally, as unobserved confounders and endogeneity may remain, for example, intrinsic factors such as motivation or access to mentorship may simultaneously influence both career trajectories and advancement outcomes. Observed geographic differences may reflect a complex combination of agglomeration effects, self-selection into large metropolitan areas, and localized job-switching patterns.

Future research could develop multi-level frameworks linking personal mobility to organizational practices, talent retention strategies, demographic disparities, and broader professional or societal outcomes. Addressing these layers would provide a more comprehensive foundation for reform in planning education, professional development, and staffing.

# 6. CONCLUSION

This study provides the first large-scale computational analysis of career mobility among U.S. planning alumni. By integrating boundaryless career theory, social capital theory, and spatial opportunity models with LLM-based annotation of 130,301 LinkedIn profiles, we identify robust evidence that multisectoral breadth, strategic geographic mobility, expanding professional networks, and a complementary blend of technical and soft skills collectively drive upward career advancement in planning.

These findings offer actionable guidance for planning educators, employers, and policymakers: curricula should embed cross-sector exposure, network-building, and flexible career navigation alongside technical instruction, while workforce development programs should reduce structural barriers to mobility and professional participation. Addressing these dimensions may support a more adaptive, connected, and equitable planning workforce.

## ACKNOWLEDGEMENTS

This manuscript is based on work supported by the National Science Foundation under Grant No. 2440023. The pedagogical research data access was supported in part by the University of Florida's self-supported online Master of Urban and Regional Planning program. Any opinions, findings, and conclusions or recommendations expressed in this material are those of the authors and do not necessarily reflect the views of the National Science Foundation or the University of Florida.

**Declaration of Use of Generative AI.** During the preparation of this manuscript, the authors used ChatGPT-4o to assist with language editing and coding. All AI-assisted content was reviewed and edited by the authors.

## REFERENCES

[1] Song, Y. (2025). Centennial reflections: A century of shaping tomorrow's cities. Journal of the American Planning Association, 91(4), 503–505. https://doi.org/10.1080/01944363.2025.2548173

[2] Where do planners work? (n.d.). American Planning Association. https://www.planning.org/choosingplanning/where/

[3] 2025 trend report for planners. (n.d.). American Planning Association. https://www.planning.org/publications/document/9304884/

[4] Kalleberg, A. L., & Mouw, T. (2018). Occupations, organizations, and intragenerational career mobility. Annual Review of Sociology, 44(1), 283–303. https://doi.org/10.1146/annurev-soc-073117-041249

[5] Sacchi, S., Kriesi, I., & Buchmann, M. (2016). Occupational mobility chains and the role of job opportunities for upward, lateral and downward mobility in Switzerland. Research in Social Stratification and Mobility, 44, 10–21. https://doi.org/10.1016/j.rssm.2015.12.001

[6] Defillippi, R. J., & Arthur, M. B. (1994). The boundaryless career: A competency-based perspective. Journal of Organizational Behavior, 15(4), 307–324. https://doi.org/10.1002/job.4030150403

[7] Rosenfeld, R. A. (1992). Job mobility and career processes. Annual Review of Sociology, 18(1), 39–61. https://doi.org/10.1146/annurev.so.18.080192.000351

[8] Arthur, M. B., & Rousseau, D. M. (2001). The boundaryless career: A new employment principle for a new organizational era. Oxford University Press.

[9] Hall, D. T. (2004). The protean career: A quarter-century journey. Journal of Vocational Behavior, 65(1), 1–13.

[10] McClure, L., Ruth, P., Dedekorkut-Howes, A., & Sloan, M. (2012). What do current planning students and recent graduates think planners do? Proceedings of the Australia & New Zealand Association of Planning Schools Conference.

[11] Guzzetta, J. D., & Bollens, S. A. (2003). Urban planners' skills and competencies: Are we different from other professions? Does context matter? Do we evolve? Journal of Planning Education and Research, 23(1), 96–106. https://doi.org/10.1177/0739456X03255426

[12] Keyes, L. M., & Ames Fischer, L. (2025). Where do I go from here? Evaluating professional development in undergraduate planning education. Journal of Planning Education and Research, 45(4), 923–935. https://doi.org/10.1177/0739456X241260609

[13] Norris-Tirrell, D., Rinella, J., & Pham, X. (2018). Examining the career trajectories of nonprofit executive leaders. Nonprofit and Voluntary Sector Quarterly, 47(1), 146–164. https://doi.org/10.1177/0899764017722023

[14] Richardson, B., Kettles, D., Mazzola, D., & Li, H. (2024). Career trajectory analysis of Fortune 500 CIOs: A LinkedIn perspective. Communications of the Association for Information Systems, 55(1), 654–679. https://doi.org/10.17705/1CAIS.05525

[15] Jacobs, E. (2025). On the up and up: The job mobility of skilled return migrants. Social Forces, 103(4), 1538–1559. https://doi.org/10.1093/sf/soae132

[16] Lee, C. C. M., Vear, A., Howard, B., & Choate, J. (2025). Tracking graduate outcomes of undergraduate physiology major students. Advances in Physiology Education, 49(2), 297–303. https://doi.org/10.1152/advan.00240.2024

[17] Tolkach, D., & Tung, V. W. S. (2019). Tracing hospitality and tourism graduates' career mobility. International Journal of Contemporary Hospitality Management, 31(10), 4170–4187. https://doi.org/10.1108/IJCHM-10-2018-0857

[18] Sun, B., Ruan, A., Peng, B., & Lu, W. (2022). Talent flow network, the life cycle of firms, and their innovations. Frontiers in Psychology, 13. https://doi.org/10.3389/fpsyg.2022.788515

[19] Gunz, H. P., & Peiperl, M. (2007). Handbook of career studies. SAGE Publications.

[20] Forrier, A., Sels, L., & Stynen, D. (2009). Career mobility at the intersection between agent and structure: A conceptual model. Journal of Occupational and Organizational Psychology, 82(4), 739–759. https://doi.org/10.1348/096317909X470933

[21] Kornblum, A., Unger, D., & Grote, G. (2018). When do employees cross boundaries? Individual and contextual determinants of career mobility. European Journal of Work and Organizational Psychology, 27(5), 657–668. https://doi.org/10.1080/1359432X.2018.1488686

[22] Lauria, M., & Long, M. (2017). Planning experience and planners' ethics. Journal of the American Planning Association, 83(2), 202–220. https://doi.org/10.1080/01944363.2017.1286946

[23] Loh, C. G., & Arroyo, R. L. (2017). Special ethical considerations for planners in private practice. Journal of the American Planning Association, 83(2), 168–179. https://doi.org/10.1080/01944363.2017.1286945

[24] Shin, E. J. (2022). Representation and wage gaps in the planning profession: A focus on gender and race/ethnicity. Journal of the American Planning Association, 88(4), 449–463. https://doi.org/10.1080/01944363.2021.1996263

[25] Doeringer, P. B., & Piore, M. J. (2020). Internal labor markets and manpower analysis. Routledge.

[26] Borck, R., & Wrede, M. (2018). Spatial and social mobility. Journal of Regional Science, 58(4), 688–704. https://doi.org/10.1111/jors.12382

[27] McCollum, D., Liu, Y., Findlay, A., Feng, Z., & Nightingale, G. (2018). Determinants of occupational mobility: The importance of place of work. Regional Studies, 52(12), 1612–1623. https://doi.org/10.1080/00343404.2018.1424993

[28] National Academies of Sciences, Engineering, and Medicine. (2025). Economic and social mobility: New directions for data, research, and policy. https://doi.org/10.17226/28456

[29] Michelangeli, A., & Türk, U. (2021). Cities as drivers of social mobility. Cities, 108, 102969. https://doi.org/10.1016/j.cities.2020.102969

[30] Giorgi, J., Plunket, A., & Starosta de Waldemar, F. (2025). Inter-regional highly skilled worker mobility and technological novelty. Research Policy, 54(1), 105113. https://doi.org/10.1016/j.respol.2024.105113

[31] Hane-Weijman, E. (2021). Skill matching and mismatching: Labour market trajectories of redundant manufacturing workers. Geografiska Annaler: Series B, Human Geography, 103(1), 21–38. https://doi.org/10.1080/04353684.2021.1884497

[32] Norton, R. K. (2005). More and better local planning: State-mandated local planning in coastal North Carolina. Journal of the American Planning Association, 71(1), 55–71. https://doi.org/10.1080/01944360508976405

[33] Granovetter, M. S. (1973). The strength of weak ties. American Journal of Sociology, 78(6), 1360–1380. https://doi.org/10.1086/225469

[34] Lin, N. (1999). Social networks and status attainment. Annual Review of Sociology, 25(1), 467–487. https://doi.org/10.1146/annurev.soc.25.1.467

[35] Cangialosi, N., Odoardi, C., Peña-Jimenez, M., & Antino, M. (2023). Diversity of social ties and employee innovation. Revista de Psicología del Trabajo y de las Organizaciones, 39(2), 65–74.

[36] Fu, J. S. (2024). Network portfolio diversity and social innovation: An egocentric approach to cross-sector partnerships. Social Networks, 78, 238–252.

[37] Putnam, R., et al. (2004). Using social capital to help integrate planning theory, research, and practice. Journal of the American Planning Association, 70(2), 142–192. https://doi.org/10.1080/01944360408976369

[38] Fugate, M., Kinicki, A. J., & Ashforth, B. E. (2004). Employability: A psycho-social construct, its dimensions, and applications. Journal of Vocational Behavior, 65(1), 14–38.

[39] Batty, M. (2019). Urban analytics defined. Environment and Planning B: Urban Analytics and City Science, 46(3), 403–405. https://doi.org/10.1177/2399808319839494

[40] Clayton, P., Goodspeed, R., Green, J., Lassiter, A., Riggs, W., & Wilson, B. (2025). More than analytics: Five approaches to educating professionals to shape today's digital cities. Journal of Planning Education and Research, 45(4), 726–732. https://doi.org/10.1177/0739456X241261372

[41] Kontokosta, C. E. (2021). Urban informatics in the science and practice of planning. Journal of Planning Education and Research, 41(4), 382–395. https://doi.org/10.1177/0739456X18793716

[42] U.S. Census Bureau. (2024). Metropolitan and micropolitan statistical areas population estimates, July 1, 2024. https://www.census.gov/data/tables/time-series/demo/popest/2020s-total-metro-and-micro-statistical-areas.html

[43] U.S. Bureau of Labor Statistics. (2023). Quarterly Census of Employment and Wages: 2023 annual data by area and by industry. https://www.bls.gov/cew/downloadable-data-files.htm

[44] Pielou, E. C. (1966). The measurement of diversity in different types of biological collections. Journal of Theoretical Biology, 13, 131–144. https://doi.org/10.1016/0022-5193(66)90013-0

[45] National Center for O*NET Development. (2025). Urban and regional planners (19-3051.00). O*NET OnLine. https://www.onetonline.org/link/summary/19-3051.00

[46] Liu, H., & Ge, Y. (2023). Job and employee embeddings: A joint deep learning approach. IEEE Transactions on Knowledge and Data Engineering, 35(7), 7056–7067. https://doi.org/10.1109/TKDE.2022.3180593

[47] Manning, C. D. (2015). Computational linguistics and deep learning. Computational Linguistics, 41(4), 701–707. https://doi.org/10.1162/COLI_a_00239

[48] Holistic Evaluation of Language Models (HELM). (n.d.). https://crfm.stanford.edu/helm/classic/latest/#/leaderboard

[49] Liu, N. F., Lin, K., Hewitt, J., Paranjape, A., Bevilacqua, M., Petroni, F., & Liang, P. (2023). Lost in the middle: How language models use long contexts (arXiv:2307.03172). https://doi.org/10.48550/arXiv.2307.03172

[50] 2025 Planners' salary and benefits survey. (n.d.). American Planning Association. https://www.planning.org/salary/2025/

[51] Barchers, C., Renski, H., & Green, J. (2025). What does the job market want from planners? Using online job descriptions to measure the demand for planning skills. Journal of the American Planning Association, 1–17. https://doi.org/10.1080/01944363.2025.2513261

[52] Bapna, S., & Funk, R. J. (n.d.). Interventions for improving professional networking for women: Experimental evidence from the IT sector. https://dx.doi.org/10.25300/MISQ/2021/15620

[53] Chang, R., Wei, X., Zhang, X., Xiong, H., & Zhu, H. (2024). How recommendation letters affect career mobility: Evidence from LinkedIn. Computers in Human Behavior, 152, 108084. https://doi.org/10.1016/j.chb.2023.108084

[54] Haff, C. W., Childers, J. T., Forbes, J. M., Lack, B. T., Jackson, G. R., & Sabesan, V. J. (2025). Gender influence on career trajectory as a shoulder and elbow surgeon. Shoulder & Elbow. https://doi.org/10.1177/17585732241310513

[55] Goldstein, H. A., Bollens, S., Feser, E., & Silver, C. (2006). An experiment in the internationalization of planning education: The NEURUS program. Journal of Planning Education and Research, 25(4), 349–363.

[56] Healey, P. (2012). The universal and the contingent: Some reflections on the transnational flow of planning ideas and practices. Planning Theory, 11(2), 188–207. https://doi.org/10.1177/1473095211419333

[57] Lauria, M., & Long, M. F. (2019). Ethical dilemmas in professional planning practice in the United States. Journal of the American Planning Association, 85(4), 393–404. https://doi.org/10.1080/01944363.2019.1627238

[58] Whittemore, A. H. (2015). Practitioners theorize, too: Reaffirming planning theory in a survey of practitioners' theories. Journal of Planning Education and Research, 35(1), 76–85. https://doi.org/10.1177/0739456X14563144

[59] Bright Data. (2024). LinkedIn profiles dataset [Dataset]. Bright Data Ltd. https://brightdata.com

[60] Planning Accreditation Board. (2025). Accredited programs (as of January 1, 2025). https://www.planningaccreditationboard.org/accredited-programs/all/

[61] Association of Collegiate Schools of Planning. (2023). Planning programs 2023–34. https://www.acsp.org/page/PlanningPrograms

[62] U.S. Social Security Administration. (n.d.). Retirement age. https://www.ssa.gov/benefits/retirement/planner/agereduction.html